\documentclass[amsmath,amssymb,11pt,superscriptaddress,reprint, preprintnumbers, notitlepage,aps,prl,twocolumn]{revtex4-1}

\pdfoutput=1 
\usepackage[utf8]{inputenc}
\usepackage[english]{babel}
\usepackage{amsmath}
\usepackage{graphicx}
\usepackage{dcolumn}
\usepackage{pbox}
\usepackage{amssymb}
\usepackage{epsfig}
\usepackage{slashed}
\usepackage{amssymb}
\usepackage{ mathrsfs }
\usepackage{color}
\usepackage[font=small]{caption}
\usepackage[font=small]{subcaption}
\usepackage{url}
\definecolor{MyDarkBlue}{rgb}{0.1, 0.1, 0.8} 
\definecolor{SBlue}{rgb}{0.2, 0.4, 0.7} 
\definecolor{MyLightBlue}{rgb}{0.22,0.51,0.9}
\definecolor{MyGreen}{rgb}{0.0, 0.5, 0.0}
\definecolor{BrickRed}{rgb}{0.8, 0.25, 0.33}
\RequirePackage{hyperref}
\hypersetup{colorlinks, citecolor=SBlue,linkcolor=MyDarkBlue, urlcolor=BrickRed}

\makeatletter

\makeatletter
\renewcommand\@makecaption[2]{%
  \par
  \vskip\abovecaptionskip
  \begingroup
  
   \small\rmfamily
    \begingroup
     \samepage
     \flushing
     \let\footnote\@footnotemark@gobble
     \@make@capt@title{#1}{#2}\par
    \endgroup
  \endgroup
  \vskip\belowcaptionskip
}
\makeatother

\DeclareUnicodeCharacter{2212}{-}
\begin{document}
\preprint{OSU-HEP-20-01}

\title{\Large 
Resolving electron and muon $g-2$  within the 2HDM
}
    
\author{\bf Sudip Jana}
\email[E-mail:]{ sudip.jana@mpi-hd.mpg.de}
\affiliation{Max-Planck-Institut f{\"u}r Kernphysik, Saupfercheckweg 1, 69117 Heidelberg, Germany}

\author{\bf Vishnu P.K.}
\email[E-mail:]{ vipadma@okstate.edu}
\affiliation{Department of Physics, Oklahoma State University, Stillwater, OK, 74078, USA}

\author{\bf Shaikh Saad}
\email[E-mail:]{ shaikh.saad@okstate.edu}
\affiliation{Department of Physics, Oklahoma State University, Stillwater, OK, 74078, USA}

\begin{abstract}
Recent precise measurement of the electron anomalous magnetic moment (AMM) adds to the  longstanding tension of the muon AMM and together strongly point towards physics beyond the Standard Model (BSM). In this work, we propose a solution to both anomalies in an economical fashion via a light scalar that emerges from a second Higgs doublet and  resides in the $\mathcal{O}(10)$-MeV to $\mathcal{O}(1)$-GeV mass range yielding the right sizes and signs for these deviations due to one-loop and two-loop dominance for the muon and the electron, respectively. A  scalar of this type is subject to a number of various experimental constraints, however, as we show,  it can remain sufficiently light by  evading all experimental bounds and has the great potential to be discovered  in the near-future low-energy experiments. The analysis provided here is equally applicable to any BSM scenario for which a light scalar is allowed to have  sizable flavor-diagonal couplings to the charged leptons. In addition to the light scalar, our theory predicts the existence of a nearly degenerate charged scalar and a pseudoscalar, which have masses of the order of the  electroweak scale.  We analyze possible ways to probe  new-physics signals at colliders and  find that this scenario can be tested at the LHC by looking at the novel process $pp \to H^\pm H^\pm jj \to l^\pm l^\pm j j + {E\!\!\!\!/}_{T}$ via same-sign pair production of charged Higgs bosons.
\end{abstract}

\maketitle
\section{Introduction}
\vspace{-0.1in}
For a charged elementary particle with half-integer intrinsic spin, the Land\'e g-factor at the tree-level has the value $g=2$. Any departure from this is called the anomalous magnetic moment (AMM)  defined by $a=(g-2)/2$. The first radiative correction to this value at the one-loop level was performed by Schwinger \cite{Schwinger:1948iu}. Our current best understanding of physics at the fundamental scale is precisely described by the Standard Model (SM) and, within this theory, contribution to $a_{SM}$ arises from loops containing Quantum Electrodynamics (QED) corrections, hadronic (QCD) processes, and electro-weak (EW)  pieces. For the electron and the muon, the QED contributions \cite{Schwinger:1948iu,Sommerfield:1957zz,Petermann:1957hs,Kinoshita:1981vs,Kinoshita:1990wp,Laporta:1996mq,Degrassi:1998es,Kinoshita:2004wi,Kinoshita:2005sm,Passera:2006gc,Kataev:2006yh,Aoyama:2007mn,Aoyama:2012wj,Aoyama:2012wk,Laporta:2017okg,Aoyama:2017uqe,Volkov:2017xaq,Volkov:2018jhy} to the AMMs, which are the most dominant corrections, have been computed up to 5-loop order   \cite{Tanabashi:2018oca,Mohr:2000ie,Czarnecki:1998nd}.  Furthermore, within the SM the accurately computed corrections from  QCD \cite{Jegerlehner:1985gq,Lynn:1985sq,Swartz:1994qz,Martin:1994we,Eidelman:1995ny,Krause:1996rf,Davier:1998si,Jegerlehner:1999hg,Jegerlehner:2003qp,Melnikov:2003xd,deTroconiz:2004yzs,Bijnens:2007pz,Davier:2007ua,Bijnens:1995xf,Hayakawa:1997rq,Knecht:2001qf,Knecht:2001qg,RamseyMusolf:2002cy,Prades:2009tw,Kataev:2012kn,Kurz:2015bia,Colangelo:2017qdm, Campanario:2019mjh} and EW \cite{Czarnecki:1995wq,Czarnecki:1995sz,Czarnecki:1996if,Czarnecki:2002nt,Heinemeyer:2004yq,Gribouk:2005ee} interactions can be important due to the current experimental precision.  

Since the electron and the muon AMMs $a_{e,\mu}$ can be measured with great precision in the experiments, and simultaneously can be computed  with outstanding accuracy within the SM, these two quantities are the most crucial observables  in particle physics. A slight deviation of these measured quantities  from the SM values will be a direct indication of physics beyond the SM (BSM). Hence, any BSM particle that couples to a lepton ($\ell=e$ and/or $\mu$), either directly or indirectly, and contributes to its AMM $a_{\ell}$ can be probed in the experiments.

In the muon sector, there has been a longstanding tension between the theoretical prediction and the value measured at the Brookhaven National Laboratory  \cite{Bennett:2006fi}, corresponding to a deviation:
\begin{align}
\Delta a_{\mu}= (2.74\pm 0.73)\times 10^{-9}.\label{mumu}
\end{align}
Since their first measurement of $\Delta a_{\mu}$ \cite{Brown:2001mga}, the discrepancy has   been slowly growing due to the reduction of  both theoretical and experimental uncertainties and has gained a lot of attention to the theory community over the last almost two decades; for reviews see e.g. Refs. \cite{Jegerlehner:2009ry,Lindner:2016bgg}.  In the near future, Fermilab's E989 Muon $g-2$ experiment \cite{Grange:2015fou} that has the precision a factor of four times better than the previous experimental measurements, is likely to publish their first result, which makes the scenario even more exciting.  Additionally, the future J-PARC experiment \cite{Abe:2019thb} developed by the E34 collaboration also aims to measure muon AMM with similar precision.  Moreover, the objective of the recently proposed MUonE experiment \cite{Abbiendi:2016xup} at CERN  is to  determine the hadronic contribution to muon AMM with a precision smaller than the theoretical uncertainty originating from leading-order QCD processes. For discussions  on possible new-physics (NP) effects on the measurements  of the MUonE experiment, see Refs. \cite{Dev:2020drf, Masiero:2020vxk}.

On the other hand, just recently an improved  measurement \cite{Parker:2018vye}  of  the  fine-structure constant  $\alpha$ using Caesium atom  points toward a deviation in the electron AMM from theoretical prediction as well:  
\begin{align}
\Delta a_{e}= -(8.7\pm 3.6)\times 10^{-13}.\label{ee}
\end{align}
Eq. \eqref{ee} corresponds to a negative $\sim 2.4 \sigma$ discrepancy for the electron, whereas Eq. \eqref{mumu} for the muon signifies a positive $\sim 3.7 \sigma$ deviation from the SM predictions. These tantalizing  disparities could play a significant role in finding clues of NP BSM.  Note however that having opposite signs of these two anomalies, along with the fact that the mass ratio of the muon to the electron is $\sim \mathcal{O}(100)$, makes it more  difficult to explain them simultaneously  within a common BSM origin.

In the literature, a few different mechanisms are proposed to take into account these deviations, e.g.,  by introducing scalar degrees of freedom \cite{Davoudiasl:2018fbb,Liu:2018xkx,Han:2018znu,Bauer:2019gfk,Cornella:2019uxs,Endo:2020mev,Haba:2020gkr,Bigaran:2020jil}, in the supersymmetric context \cite{Badziak:2019gaf,Endo:2019bcj,Dong:2019iaf,Dutta:2018fge}, utilizing vector-like fermions \cite{Hiller:2019mou,Crivellin:2019mvj,Crivellin:2018qmi}, in models with gauge-extensions  \cite{Abdullah:2019ofw, CarcamoHernandez:2020pxw,CarcamoHernandez:2019ydc}, and considering non-local QED effects \cite{He:2019uvu}.  Whereas most of the constructions are effective theories or require the presence of additional fermionic states, in this work, we propose  a simple ultraviolet (UV) complete model without extending the gauge sector of the SM and without introducing  BSM fermionic states. In our framework, the observed disparities of the lepton AMMs given in Eqs. \eqref{mumu}-\eqref{ee} have a common origin, and
proper explanation of both these anomalies relies on the existence of a new light scalar degree of freedom that resides in the $\mathcal{O}(10)$-MeV to $\mathcal{O}(1)$-GeV mass range. NP around this low-energy regime  is very interesting and has the potential to be probed in the ongoing, as well as in the upcoming experiments.  As we will show, such a light scalar, even though subject to a number of various experimental constraints, can simultaneously incorporate the deviations observed in the muon and the electron AMMs.

Our UV-complete theory is the well-motivated two-Higgs-doublet-model (2HDM) \cite{Lee:1973iz, Branco:2011iw}, which is one of the simplest extensions of the SM.  A variety of theories beyond the SM naturally contain a second Higgs doublet, such as supersymmetric theories \cite{Haber:1984rc}, axion models to solve the strong CP-problem \cite{Peccei:1977hh, Kim:1986ax}, left-right symmetric models \cite{Senjanovic:1975rk}, and more.  In this theory, in addition to the SM Higgs $h$, there exist one CP-even $H$, one CP-odd $A$, and a charged $H^+$ physical scalars.  We show that the new CP-even state can remain significantly light ($m_{H}\ll m_h, m_{A}, m_{H^+}$) evading all experimental constraints and contribute to both $a_{\mu}$ and $a_{e}$ to the right amounts. Even though the corrections to each of the AMMs arise from $H$ mediated one-loop and two-loop processes, a positive one-loop quantum correction  dominates for the muon AMM, whereas  the required contribution to the electron AMM originates primarily from a two-loop diagram that has a sign ambiguity.  
For elaborated discussions on loop-mediated contributions to lepton AMMs via scalars see e.g. Refs.  \cite{Giudice:2012ms, Marciano:2016yhf}. For explanations of only the muon $g-2$ within the 2HDM see e.g. Refs. 
\cite{Haber:1978jt, Krawczyk:1996sm, Nie:1998dg, Dedes:2001nx, Iltan:2001nk, Krawczyk:2001pe, Wu:2001vq, Gunion:2008dg, Cao:2009as, Broggio:2014mna, Chun:2016hzs, Wang:2014sda, Abe:2015oca, Crivellin:2015hha, Chun:2015hsa, Han:2015yys, Ilisie:2015tra, Cherchiglia:2016eui, Abe:2017jqo, Cherchiglia:2017uwv, Keus:2017ioh, Li:2018aov, Wang:2018hnw, Chun:2019oix, Iguro:2019sly}.  It should be stressed that
the analysis provided in this work is equally applicable to any BSM scenario for which  the effective theory consists of a light  CP-even state \cite{Bertuzzo:2018ftf} having sizable flavor-diagonal couplings to charged leptons and negligible mixing with the SM Higgs.    

In the next section we introduce the proposed model, then in Sec. \ref{SECa} we summarize experimental constraints relevant to our study and  present detailed results, and finally, we conclude in Sec. \ref{SECb}.

\section{Model}\label{SEC-1}
\textbf{Scalar sector:}-- 
In our proposed model, the SM containing a Higgs doublet $\Phi_1$ is extended by a second Higgs doublet $\Phi_2$, each carrying hypercharge=1/2.  Both the Higgs doublets can acquire vacuum expectation values (VEVs) $\langle \Phi_i\rangle= v_i$, such that $v_iv_j\delta_{ij}=v^2=(246$ GeV$)^2$. This introduces a parameter in the theory defined as: $\tan\beta=v_2/v_1$. However, one can choose a particularly convenient rotated basis in which only one neutral Higgs has a nonzero vacuum expectation value.   The most general scalar potential of 2HDM written in this so-called Higgs-basis is given by \cite{Wu:1994ja, Davidson:2005cw, Branco:2011iw,  Babu:2018uik}:
\begin{align}
&V= m_{11}^2H_1^{\dagger}H_1+m_{22}^2H_2^{\dagger}H_2
-\{m_{12}^2H_1^{\dagger}H_2+{\rm h.c.}\} 
\nonumber\\ &
+\frac{\lambda_1}{2}(H_1^{\dagger}H_1)^2
+\frac{\lambda_2}{2}(H_2^{\dagger}H_2)^2
+\lambda_3(H_1^{\dagger}H_1)(H_2^{\dagger}H_2)
\nonumber\\ &
+\lambda_4(H_1^{\dagger}H_2)(H_2^{\dagger}H_1)
+\left\{\frac{\lambda_5}{2}(H_1^{\dagger}H_2)^2+{\rm h.c.}\right\}
\nonumber\\ &
+\left\{
\big[\lambda_6(H_1^{\dagger}H_1)
+\lambda_7(H_2^{\dagger}H_2)\big]
H_1^{\dagger}H_2+{\rm h.c.}\right\}.
\end{align}
Here  $m_{12}^2$, and $\lambda_{5,6,7}$  can be complex in general, whereas the rest of the parameters are real. We work in the CP-conserving limit and take all the parameters to be real.  The Higgs-basis and the original basis are related by the following transformations: 
\begin{align}
&H_1=\cos\beta \;\Phi_1+\sin\beta\; \Phi_2,
\\ 
&H_2=-\sin\beta \;\Phi_1+\cos\beta\; \Phi_2.
\end{align}
Note that in this basis, only $H_1$ has non-zero VEV \cite{Babu:2018uik}, and these fields can be parametrized as: 
\begin{align}
H_1=\begin{pmatrix}
G^+\\
\frac{v+H_1^0+i G^0}{\sqrt{2}}
\end{pmatrix},\;
H_2=\begin{pmatrix}
H^+\\
\frac{H_2^0+i A^0}{\sqrt{2}}
\end{pmatrix}.
\end{align}
Here $G^+$ and $G^0$ are the Goldstone bosons eaten up by the gauge bosons after the EW  symmetry is broken.  Furthermore, $H^0_{1,2}$ are the CP-even neutral  and $A^0$ is the CP-odd neutral scalars. The mass eigenstates of the CP-even neutral scalars are as follows \cite{Babu:2018uik}:
\begin{align}
&h=\cos(\alpha-\beta)\;H^0_1 +\sin(\alpha-\beta)\;H^0_2,\\
&H=-\sin(\alpha-\beta)\;H^0_1 +\cos(\alpha-\beta)\;H^0_2.
\end{align}
Here the corresponding mixing angle is defined as \cite{Babu:2018uik}:
\begin{align}
\sin 2(\alpha-\beta)=\frac{2 v^2 \lambda_6}{m^2_H-m^2_h}.    
\end{align}
In our study, we work in the alignment limit \cite{Bernon:2015qea, Branco:2011iw, Babu:2018uik, Dev:2014yca}, which by following the above definitions corresponds to $\alpha\approx \beta$ \cite{Babu:2018uik}. 
In this limit, $H_1^0\approx h$ is the SM Higgs and almost decouples from the other CP-even state $H_2^0\approx H$. Then the masses of all the physical scalars in this theory are given by \cite{Babu:2018uik}:
\begin{align}
&m^2_h= \lambda_1v^2,\;
m^2_H=m^2_{22}+\frac{v^2}{2}(\lambda_3+\lambda_4+\lambda_5),
\\
&m^2_A=m^2_H-v^2 \lambda_5,\;
m^2_{H^\pm}=m^2_H-\frac{v^2}{2}(\lambda_4+\lambda_5).
\end{align}

As aforementioned, we are interested in the scenario with a mass hierarchy of the form: $m^2_{H} \ll m^2_{H^{+}},  m^2_{A}$. With this choice, the EW precision measurements put restrictions on the mass splitting between $H^+$ and $A^0$ states, hence, for  simplicity, we take them to be degenerate, $m^2_{H^{\pm}}= m^2_{A^0}$.  This demands, $\lambda_4=\lambda_5 (\equiv \lambda)$, and consequently one finds: $m^2_{H^+}=m^2_A=-v^2 \lambda$ (here we have neglected the small mass of $H$ scalar). From this, it is evident that masses of the heavy scalars $H^+$ and $A$ cannot be made arbitrarily large. Perturbatively of the couplings $|\lambda| \lesssim 2$ (or $\sqrt{4\pi}$) provides an upper bound on the mass of the heavy states $m_{H^+}=m_A \lesssim 350$ (or 460)  GeV, as long as $m_H\approx 0$.

On the other hand,  a lower bound on the charged Higgs mass utilizing LEP constraints in our scenario is found to be $m_{H^{+}}\geq 110$ GeV as will be discussed later in the text. For the simplicity of our analysis, we fix its mass to be $110$ GeV for the  rest of this work that corresponds to the case $\Delta m\equiv m_{H}-m_{H^{\pm}}=-110$ GeV. This bound can be then translated to $\lambda=\Delta m^2/v^2=-0.199$, which  essentially remains the same for a wide range of mass 0 GeV $\leq m_{H} \leq$ 1 GeV, but can be significantly different in the larger mass region. 

\textbf{Yukawa sector:}-- 
The Yukawa coupling of this theory is given by \cite{Wu:1994ja, Davidson:2005cw, Branco:2011iw}:
\begin{align}
-&\mathcal{L}_Y\supset  
\sqrt{2}({Y^{(1)}_{k,ij}}\Phi_1+{Y^{(2)}_{k,ij}}\Phi_2){\overline{k}_L}_i{k_R}_j + h.c. \label{YUK}
\end{align}
Here for quarks $k_L=Q_L, k_R=u_R, d_R$, for leptons $k_L=L_L, k_R=\ell_R$, and  in the up-quark sector  $\Phi\to i\tau_2\Phi^{\ast}$ must be made. In the Higgs-basis the Lagrangian has the same form as that of  Eq. \eqref{YUK}  with the replacements: $\Phi_i\to H_i$, and $\{Y^{(1)}_k, Y^{(2)}_k\} \to \{\widetilde{Y}_k, \overline{Y}_k\}$, where we have defined: $\widetilde{Y}_k= Y^{(1)}_k\cos\beta + Y^{(2)}_k\sin\beta$ and $\overline{Y}_k= -Y^{(1)}_k\sin\beta + Y^{(2)}_k\cos\beta$.  Note that $\widetilde{Y}_k$ and  $\overline{Y}_k$ are independent  $3\times 3$ Yukawa coupling matrices.  Since in the Higgs-basis only $H_1$ acquires a VEV, the masses of the fermions are entirely coming from $\widetilde{Y}_k$ Yukawa couplings that follow the relations $\widetilde{Y}_k=\frac{\sqrt{2}}{v} M_k$,  whereas  $\overline{Y}_k$ are free parameters. We work in a basis, where the mass matrices are real and diagonal. In this chosen basis, the second set of Yukawa coupling matrices, $\overline{Y}_k$ are in general arbitrary non-diagonal  matrices and we denote these rotated matrices by $Y_k$. However, $Y_k$ are subject to stringent phenomenological constraints, since they mediate dangerous flavor violating processes. In the quark sector, even if one starts with diagonal $\overline{Y}_k$, off-diagonal entries reappear in $Y_k$ matrices due to non-vanishing CKM entries.  This is why  we assume all entries in both the up-type and down-type quarks to be sufficiently small $\overline{Y}_{u,ij}, \overline{Y}_{d,ij}\ll 1$, and focus only on the lepton sector.  Following the above discussions, the  Yukawa interactions of the leptons with the physical scalars are then given by: 
\begin{align}
-&\mathcal{L}_Y\supset 
\left[  
{Y^{H^0}_{\ell,ij}}  H^0
+i\;Y^{A^0}_{\ell,ij} A^0
\right] {\overline{\ell}_L}_i{\ell_R}_j
\nonumber \\ &
+Y^{H^+}_{\ell,ij} {\overline{\nu}_L}_i{\ell_R}_j H^+\sqrt{2} +h.c.,
\end{align}
here, $Y_{\ell}^{H^0}=Y_{\ell}^{A^0}=Y_{\ell}^{H^+}=Y_{\ell}$.  For this lepton Yukawa matrix, we assume a texture of the form: $Y_{\ell}= diag(y_e, y_{\mu}, y_{\tau})$, where, couplings $y_{\ell}$ are uncorrelated to the masses of the leptons and  we take them to be real. This choice of Yukawa texture is taken purely due to phenomenological considerations to avoid dangerous flavor violating processes.

\begin{widetext}
\begin{figure*}[th!]
\includegraphics[width=0.28\textwidth]{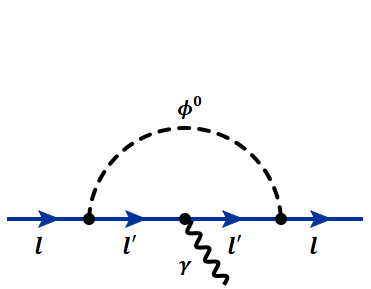}
\hspace{2cm}
\includegraphics[width=0.28\textwidth]{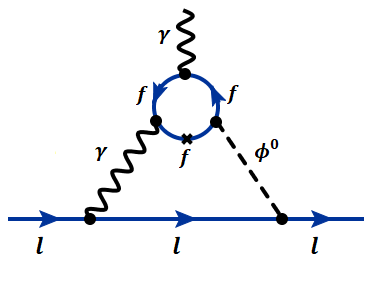}
\caption{One-loop (left) and two-loop (right) contributions to lepton AMMs arising from beyond-SM neutral scalars. The one-loop contribution due to the charged scalar is not presented here. 
For our choice of diagonal Yukawa couplings, the only term that contributes for the one-loop diagram corresponds to $\ell' =\ell$.} \label{12loop}
\end{figure*}
\end{widetext}

\textbf{Lepton anomalous magnetic moments:}--  
We remind the readers that the SM Higgs does not mix with the BSM states. Consequently the contributions of $h$ via the one-loop and two-loop diagrams of Fig. \ref{12loop} remain the same as that of SM, which is already a part of $a^{\text{SM}}_{\ell}$. Now we compute all possible BSM contributions to lepton AMMs ($\Delta a_{\ell}$) within our framework.  We first derive the one-loop contributions as shown in  Fig. \ref{12loop} (diagram on the left) arising from the charged, CP-even,  and CP-odd scalars  which are given by   \cite{Leveille:1977rc}:  
\begin{align}
&\Delta a^{H^+}_{1,\ell}=
\frac{-Q_{H^+}\left(Y^{H^+}_{\ell}\right)^2}{4\pi^2} F_{H^+}[z_{H^+}],
\\ 
&\Delta a^{\phi^0}_{1,\ell}=
\frac{-1}{8\pi^2} \sum_{\phi^0=}^{H,A} 
Q_{\ell}
\left(Y^{\phi^0}_{\ell}\right)^2 F_{\phi^0}[z_{\phi^0}], 
\\
&z_{H^+}= \frac{m_{H^+}}{m_{\ell}}, z_{\phi^0}= \frac{m_{\phi^0}}{m_{\ell}},
\\
&F_{H^+}[z_{H^+}]= \int_0^1 dx \frac{x^2(x-1)}{x^2+x(z^2_{H^+}-1)},
\\
&F_{\phi^0}[z_{\phi^0}]=
\int_0^1 dx \frac{x^2(1-x\pm  1)}{x^2+z^2_{\phi^0}(1-x)}.
\end{align}
 In deriving the above formulas, we have adopted the scenario of diagonal Yukawa couplings.
Moreover, in the $F_{\phi^0}$ formula,  $+$ and $-$ corresponds to the cases  $\phi^0=H$ and $\phi^0=A$, respectively. 

Within our set-up, the neutral scalars with the help of fermion loops can contribute to lepton AMMs  via a two-loop Barr-Zee diagram \cite{Bjorken:1977vt, Barr:1990vd, Barr:1990vd} as shown in Fig. \ref{12loop} (diagram on the right). We further derive these two-loop  contributions to $\Delta a_{e,\mu}$ and find these corrections to be: 
\begin{align}
&\Delta a^{\phi^0}_{2,\ell}=
\frac{\alpha}{8\pi^3}m_{\ell}Y^{\phi^0}_{\ell}
\;\sum_{f}\sum_{\phi^0=}^{H,A} \frac{N^c_fQ^2_fY^{\phi^0}_f}{m_f}F_{\phi^0}\left[ \frac{m^2_f}{m^2_{\phi^0}}  \right],   \label{eqq}   
\\
&F_{\phi^0}\left[ z_{\phi^0}  \right]=z_{\phi^0}\int_{0}^{1}dx \frac{w_{\phi^0}}{x(1-x)-z_{\phi^0}}\ln{\frac{x(1-x)}{z_{\phi^0}}},
\\
&w_{H}=2x(1-x)-1,\;\; w_{A}=1.
\end{align}
In Eq. \eqref{eqq},  the sum over the internal fermions is taken over $f=e, \mu, \tau$.

Note that in the two-loop diagram shown in Fig. \ref{12loop}, the fermions $f$ running inside the loop can be replaced by charged scalar $H^{\pm}$. Contribution of this type originates only from the $\lambda_7$ term in the scalar potential, and this quartic coupling plays no role in giving masses to the scalars. In our analysis we take this coupling to be small just for simplicity, and consequently do not include the diagram involving charged scalar loop.  It is to be mentioned that adding this contribution  to $\Delta a_{\ell}$ will not change the results of this work, since $m_H^{\pm} \gg m_H$.

\begin{widetext}
\begin{figure*}[th!]
\includegraphics[width=0.41\textwidth]{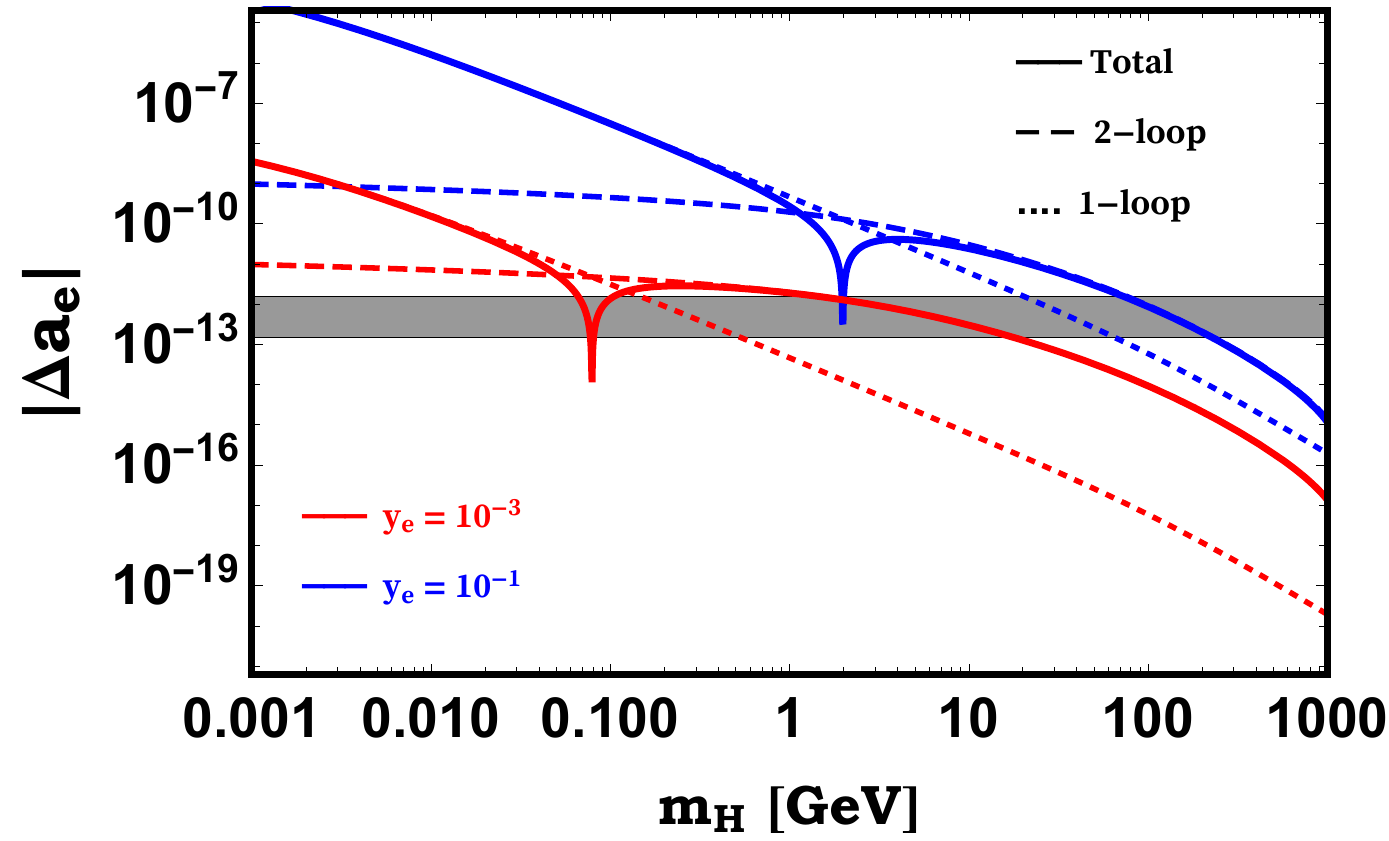}
\hspace{0.5cm}
\includegraphics[width=0.41\textwidth]{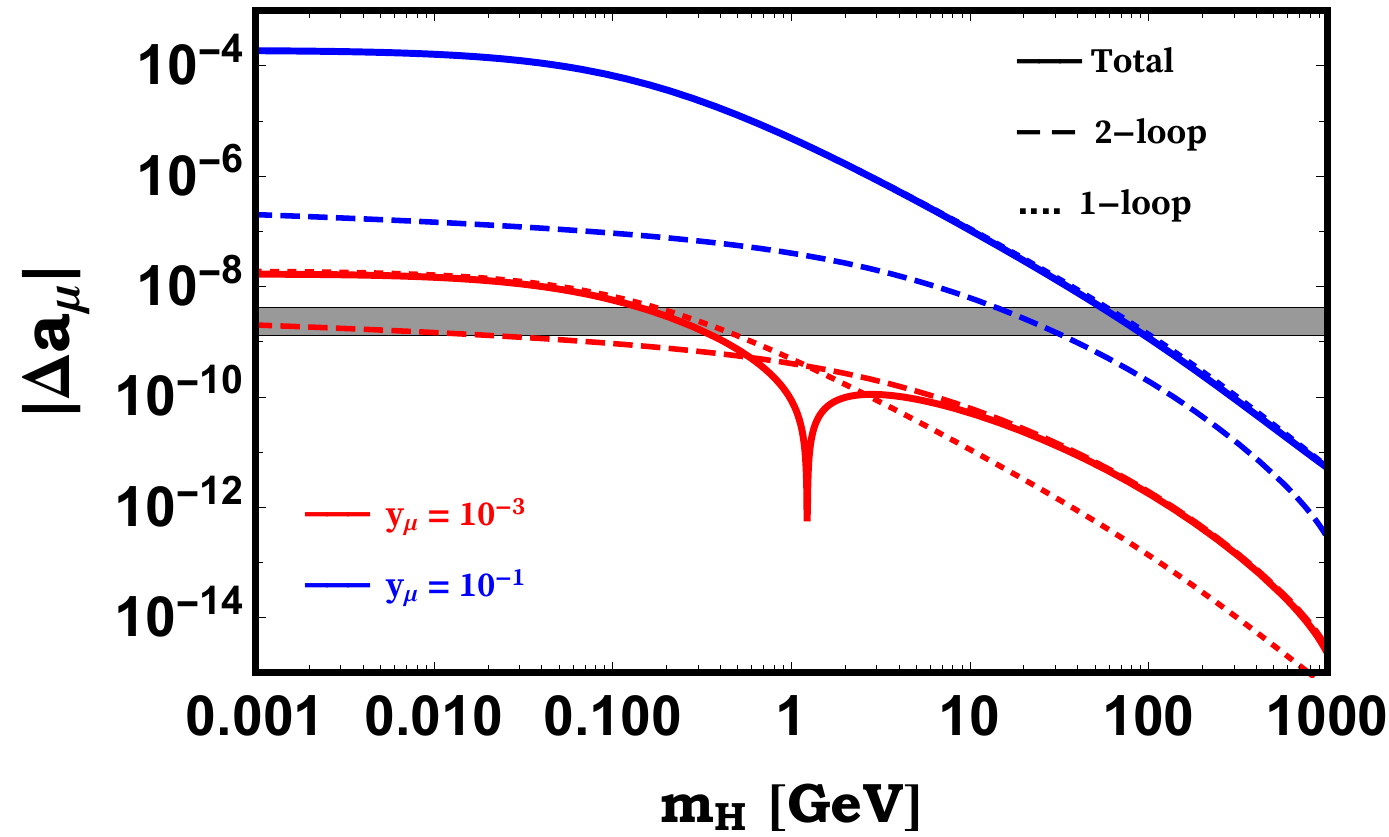}
\caption{
Magnitudes of one-loop (dotted) and two-loop (dashed) contributions to $\Delta a_{\ell}$ ($\ell=e, \mu$) by the BSM states of the theory. Solid lines represent the total magnitude of $\Delta a_{\ell}$ by assuming negative   two-loop contributions.  The horizontal bands indicate the experimental $2\sigma $ allowed region.} \label{g2e}
\end{figure*}
\end{widetext}

As aforementioned, we are interested in an interesting regime of the 2HDM where the CP-even state $H$, emerging from the second Higgs doublet remains sufficiently light compared to its partners. In our scenario, a mass splitting of this type is essential for concurrent explanation of $\Delta a_{\mu}$ and $\Delta a_{e}$. As will be apparent from the detailed analysis performed in the next section, the experimentally allowed mass window is $\mathcal{O}(10)$-MeV to $\mathcal{O}(1)$-GeV for the light scalar. In this scheme, only the contribution of the light state to the lepton AMMs is significant, since our case corresponds to $m_{H^+}=m_A \gg m_H$. Here we investigate the viability of attaining right sizes and signs for both the deviations observed in $g_{\mu}-2$ and $g_{e}-2$ measurements via CP-even scalar $H$. For completeness we have also included the contributions from $H^+$ and $A$ that can provide sizable corrections at the higher mass regime. To get an understanding of the relative magnitudes, in Fig.~\ref{g2e},  we show both the one-loop (dotted line) and the two-loop (dashed line) contributions to AMMs for two different values of the Yukawa couplings as a function of its mass $m_{H}$.  The solid lines correspond to overall contributions to $|\Delta a_{\mu,e}|$, and the horizontal gray bands indicate  the experimental measurements within their $2\sigma$ values.  From  Fig.~\ref{g2e}, one finds that within the mass range under consideration,  the positive one-loop contribution is the primary source of $\Delta a_{\mu}$,  whereas the two-loop correction with a negative sign must dominate over the positive one-loop correction to  $\Delta a_e$, to properly take into account the observed data given in Eqs. \eqref{mumu}-\eqref{ee}. 

In making these plots as well as for the rest of the analysis, we  fix the tau Yukawa coupling to be $y_{\tau}= 0.1$, which is allowed by the experiment data to be discussed later in the text. From the above analysis, it is clear that the choice of $y_{\tau}$ plays significant role in explaining  $\Delta a_{e}$ data.  The two-loop contribution to $\Delta a_{e}$  is directly proportional to the product of the  Yukawa couplings $\Delta a_{e} \propto y_e y_{\tau}$. As a result, a choice of  smaller values of $y_{\tau}$ demands larger values of $y_{e}$ to compensate the decrease in tau Yukawa coupling. As will be apparent from our detailed analysis, taking a value of $y_{\tau}$ ($y_{e}$), for example one order smaller (larger) than the above-mentioned choice will rule out almost the entire parameter space to accommodate $\Delta a_{e}$ within its $1\sigma$ measured value (see Fig. \ref{moneyplot}). Additionally, since the Barr-Zee diagram provides a negative contribution (apart from the sign of the Yukawa couplings) to AMM for a neutral CP-even scalar, the sign of the product of the Yukawa couplings must be positive.

\section{Experimental constraints and future prospects}\label{SECa}
As demonstrated above, a relatively light scalar compared to the EW scale can naturally explain the observed deviations of both the electron and the muon AMMs.  However, a light scalar of mass $m_H<$ TeV,  having sizable couplings to the SM charged leptons is subject to diverse experimental constraints. In search of finding the allowed parameter space of our theory, in the following, we summarize and analyze in great detail all the relevant experimental constraints, and  discuss the feasibility as well as, testability of this theory.

\begin{figure*}[t!]
$$
\includegraphics[width=0.48\textwidth]{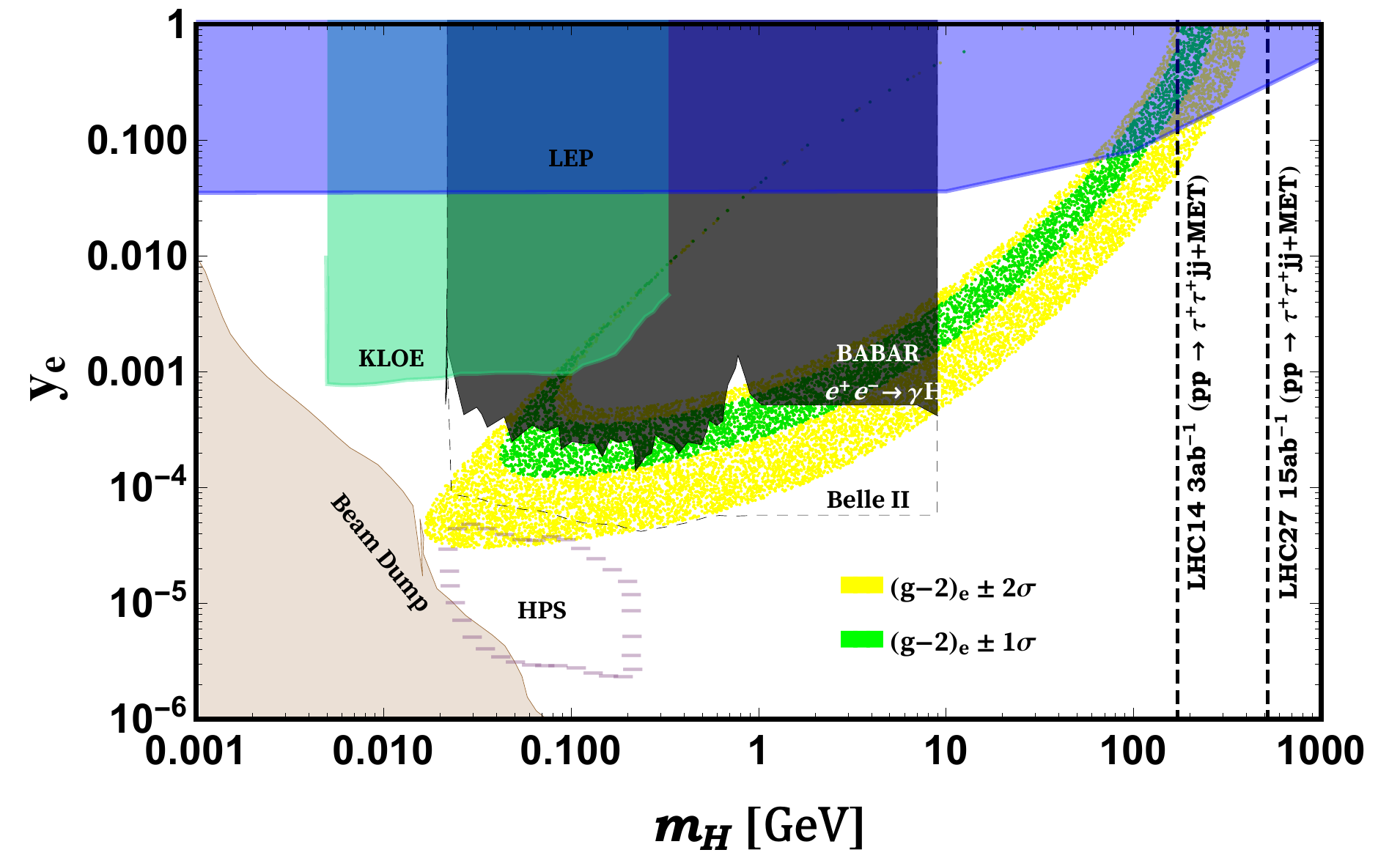}
\includegraphics[width=0.48\textwidth]{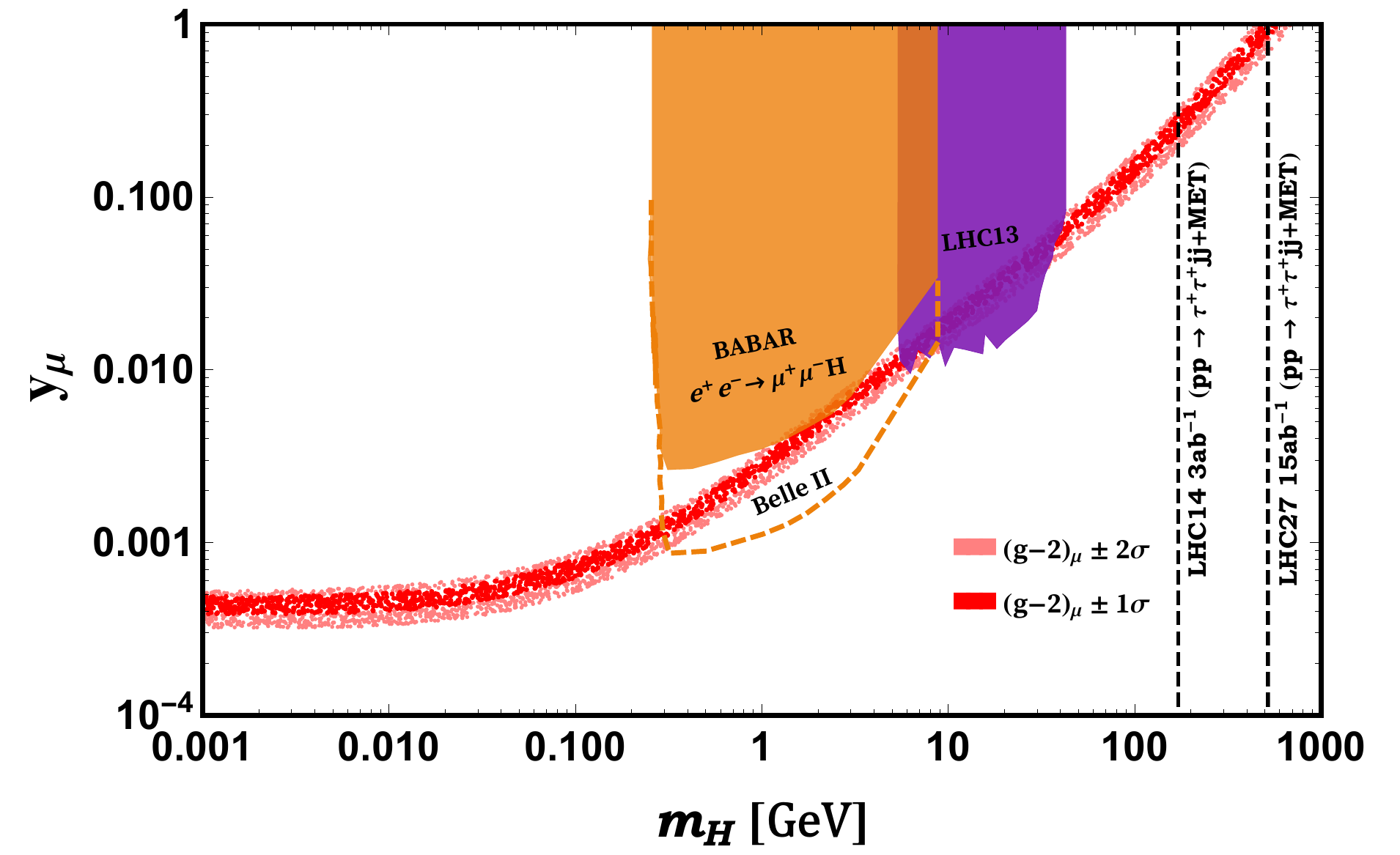}
$$
\caption{The parameter space in Yukawa coupling ($y_l$, whrere $l$= $e$ or $\mu$) vs mass ($m_H$) plane consistent with both the electron and muon AMMs. The green (red) and yellow (pink) regions represent the experimental 1$\sigma$ and 2$\sigma$ bands for the electron (muon) AMM $\Delta a_e$ ($\Delta a_{\mu}$). The color shaded regions with solid boundary denote the excluded parameter space by current experiments: brown region from the  electron beam-dump experiments \cite{Bjorken:1988as, Riordan:1987aw, Davier:1989wz}; cyan and light black regions from the dark-photon searches through $e^+e^- \rightarrow \gamma H$ process at KLOE \cite{Anastasi:2015qla} and BaBar \cite{Lees:2014xha} respectively; light brown region from the  $e^+e^- \rightarrow \mu^+\mu^- H$ searches at BaBar \cite{TheBABAR:2016rlg}; blue shaded region from LEP~\cite{LEP:2003aa}; purple region from CMS~\cite{Sirunyan:2018nnz}. In this plot, we also present the projected sensitivities from several proposed experiments: heavy-photon searches (HPS) from JLab experiment (dashed purple line) ~\cite{Battaglieri:2014hga}; dark-photon searches through $e^+e^- \rightarrow \gamma H$ process and  $e^+e^- \rightarrow \mu^+\mu^- H$ process from Belle-II (dashed black and dashed light brown lines respectively)  \cite{Abe:2010gxa,Kou:2018nap}. The projected sensitivities for the signal   $pp \to H^\pm H^\pm jj \to \tau^\pm \tau^\pm j j +$ ${E\!\!\!\!/}_{T}$  at the LHC for centre of mass energy 14 TeV with integrated luminosity $\mathcal{L}=3$ ab$^{-1}$ and also for the centre of mass energy 27 TeV with integrated luminosity $\mathcal{L}=15$ ab$^{-1}$ are shown by black dashed vertical lines.
 The $y_e$ coupling is independently constrained from electron beam-dump experiments \cite{Bjorken:1988as, Riordan:1987aw, Davier:1989wz}, the dark-photon searches through $e^+e^- \rightarrow \gamma H$ process at KLOE \cite{Anastasi:2015qla}, BaBar \cite{Lees:2014xha} and LEP~\cite{LEP:2003aa} experiment;  whereas the $y_\mu$ coupling is constrained  from the  $e^+e^- \rightarrow \mu^+\mu^- H$ searches at BaBar \cite{TheBABAR:2016rlg}  and LHC~\cite{Sirunyan:2018nnz} experiments.  } \label{moneyplot}
\end{figure*}

\textbf{Fixed-target Experiments:}-- 
Electron beam-dump experiments \cite{Bjorken:1988as, Riordan:1987aw, Davier:1989wz} can probe light scalars that have coupling with the electrons. In these experiments light scalars can be produced via Bremsstrahlung-like processes: $e + N \rightarrow e + N + H$. For a scalar of mass $m_H<2m_{\mu}$, after traveling macroscopic distances, it would  decay back to electron pairs. The lack of such events at the electron beam-dump experiments provide stringent constraints \cite{Liu:2016qwd,Batell:2016ove} on the mass of the light scalar and its corresponding couplings to the electrons, which is depicted in the brown-shaded exclusion region in Fig.~\ref{moneyplot}.

Another low-energy fixed-target experiment, HPS at the JLab \cite{Battaglieri:2014hga} is designed to search for heavy-photons. Displaced decays of scalars that are produced via their couplings to electrons can be detected in this experiment within a few cm from the target \cite{Batell:2016ove}. The HPS projection for a light scalar that couples to the electron is plotted as a dashed-purple line in  Fig.~\ref{moneyplot}.

\textbf{Dark Photon searches:}-- 
There are several experiments that search for the presence of dark-photons and their null observations  can be translated to provide stringent constraints on the allowed parameter space of light scalars.   
KLOE collaboration \cite{Anastasi:2015qla} searches for the dark-photons $A_{d}$ through the process: $e^+e^- \rightarrow \gamma A_{d}$,  with $A_{d} \rightarrow e^+e^-$. The lack of such signals at this  experiment can be used to set constraints on the light scalars  \cite{Alves:2017avw} that have coupling with the electrons, which is indicated by cyan-shaded region in  Fig.~\ref{moneyplot}.

Through a similar process, the BaBar
collaboration \cite{Lees:2014xha} also searches for the dark-photons with $A_{d} \rightarrow \ell^+\ell^-$. 
By recasting the results from BaBar,  Ref. \cite{Knapen:2017xzo} provides exclusion regions in the light scalar mass and Yukawa coupling plane, which is depicted by a  light-black shaded region in  Fig.~\ref{moneyplot}. The dashed black line below this region represents the projected sensitivity from the Belle-II experiment \cite{Abe:2010gxa,Kou:2018nap} for a similar process \cite{Batell:2016ove}.
For a scalar mass $m_{H}>200$ MeV the dark-boson searches at the BaBar \cite{TheBABAR:2016rlg} can be used to impose limits on $H \mu^+\mu^-$ coupling via $e^+  e^- \rightarrow \mu^+  \mu^- H$ process \cite{Batell:2016ove,Batell:2017kty}. We recast this result for our scenario, which is  shown as light brown shaded region in  Fig.~\ref{moneyplot}. The corresponding projected sensitivity from Belle-II experiment \cite{Kou:2018nap,Batell:2016ove} is also presented by a dashed brown line.

\textbf{Rare $Z$-decay constraints:}--
Exotic $Z$ decay of the type $Z \rightarrow 4\mu$ has been searched by both the ATLAS \cite{Aad:2014wra} and the CMS \cite{Sirunyan:2018nnz} collaborations at the LHC  with 7 TeV, as well as 8 TeV data. The LHC  results have been interpreted as constraints on the process  $Z \rightarrow \mu^{+}\mu^{-}H $, with $H \rightarrow \mu^{+}\mu^{-}$  by Ref. \cite{Batell:2017kty}. We recast the LHC results for our model, which is plotted as a purple region in  Fig.~\ref{moneyplot}.

\textbf{LEP and LHC constraints:}--
Here, we discuss the existing collider constraints on the neutral and charged scalars relevant for our set-up. Collisions of electron-positron at center-of-mass energies above the Z-boson mass are carried-out   at LEP experiment~\cite{LEP:2003aa}, which impose stringent constraints on contact interactions involving $e^+e^-\to f\overline{f}$ processes. If a neutral scalar ($\phi^0=H, A$) is heavy enough, integrating it out leads to a $d=6$  effective operator to describe the associated contact interactions.  LEP constraints are then directly translated into the lower bounds on the mass of the scalar for a given Yukawa coupling. The most constraining process is the one with electrons  in the final states and the associated bound is found to be $m_{\phi^0}/|y_e|>1.99$ TeV \cite{Babu:2019mfe}. However, if the neutral scalar is light, the aforementioned bound is no longer applicable. To properly incorporate such a  scenario, we implement our model file in FeynRules package~\cite{Christensen:2008py}  and compute the cross-section of the process $e^+e^-\to f\overline{f}$ using MadGraph5 event generator ~\cite{Alwall:2014hca}. The generated data set is then compared with the measured cross-sections ~\cite{LEP:2003aa, Abbiendi:2003dh} to find the limits on the  mass $m_{\phi^0}$ as a function of its Yukawa couplings.  The obtained LEP bounds for our model is then projected in Fig. \ref{moneyplot} in  blue-shaded region. As far as the LHC bounds, most of the searches for heavy neutral scalars are done in the context of either MSSM or generic 2HDM, which are not directly applicable in our scenario since,  $\phi^0$ has negligible couplings to quarks, and therefore, cannot be produced via gluon fusion. However,
LHC bounds on neutral scalars come out to be weaker than the LEP bounds as discussed above due to its leptophilic nature. 

Even though the charged scalars, $H^{\pm}$ do not couple to the quarks, they can still be    pair-produced through $s$-channel Drell-Yan process mediated by $Z$ or $\gamma$ at LEP. In our model, each charged scalar produced, will then decay into $\ell+\nu_{\ell}$. These leptonic final states exactly mimic slepton searches in supersymmetric models, and we use the associated LEP limits and recast these results  for our scenario, which provides a lower bound for its mass $m_{H^{\pm}}\geq 110$ GeV. The collider constraints of this type of leptophilic charged scalars are analyzed and discussed in detail in Ref.  \cite{Babu:2019mfe}.  At the LHC, the charged scalars can also be pair produced via Drell-Yan process followed by  leptonic decays $H^\pm\to \ell \nu$. Such a  leptophilic-like  charged scalar will be constrained from the LHC searches  by processes involving the left-handed  selectrons/smuons/staus \cite{Sirunyan:2018vig,Aad:2014yka, Sirunyan:2018nwe} $pp\to \widetilde{\ell}^+_L\widetilde{\ell}^-_L\to \ell^+_L\widetilde{\chi}^0\ell^-_L\widetilde{\chi}^0$, which will mimic the similar   final states $\ell^+\nu\ell^-\nu$ from $H^+H^-$ decays in the massless neutralino limit. We adapt the $\sqrt s=13$ TeV CMS selectron search~\cite{Sirunyan:2018nwe} limit and the current  limits~\cite{Sirunyan:2018vig} on stau searches, and  translate into a bound on the charged scalar mass. It is quite evident that the LHC limits can be evaded by going to larger ${\rm BR}_{\tau\nu}\gtrsim 0.4$, which is achieved in our scenario by choosing an appropriate Yukawa coupling $y_{ \tau}\sim$ 0.1.  The  LHC searches do not put any stronger bound on the mass of the leptophilic charged scalar due to its tau-philic (mostly) nature. 

It is quite important to mention that there will not be any significant constraints from Higgs observables since we are considering a scenario with almost no mixing between SM Higgs, $h$ and the other CP-even scalar, $H$. However, there would still be a coupling between the
SM Higgs and a pair of the new neutral scalar ($H$). This would imply that the SM like Higgs should have a decay to these light scalar pairs and each of these light scalar will further decay into two charged leptons. This four lepton final state signature will be similar to the $h \to ZZ^* \to l^+ l^+ l^- l^-$  except the fact that dilepton invariant mass can be reconstructed at the  light resonance instead of $Z-$ boson mass. However, for simplicity, the relevant combination of quartic couplings between the SM Higgs ($h$) and the light scalar $H$ is chosen to be small to avoid this constraint. The part of the scalar potential that contains this vertex is as follows:
\begin{align}
V\supset v\;h\;H^2 \left( \lambda+\frac{\lambda_3}{2} \right).   
\end{align}
The above-mentioned goal can be readily achieved by assuming $\lambda_3\approx -2 \lambda$. This choice is completely consistent and in this limit, the mass of $H$ is entirely determined by the free parameter $m_{22}$, whereas masses of $A$ and $H^{\pm}$ remain unaltered.  

Moreover, it is quite interesting to mention that a light neutral scalar in the mass range of (10 MeV - 1 GeV) could be probed via this Higgs-portal coupling looking at 4-lepton resonant search for the SM Higgs boson. This is a smoking gun signal of our model. The investigation of this effect is  beyond the scope of this paper and shall be presented in future work.

 \begin{figure}[t!]
\includegraphics[width=0.46\textwidth]{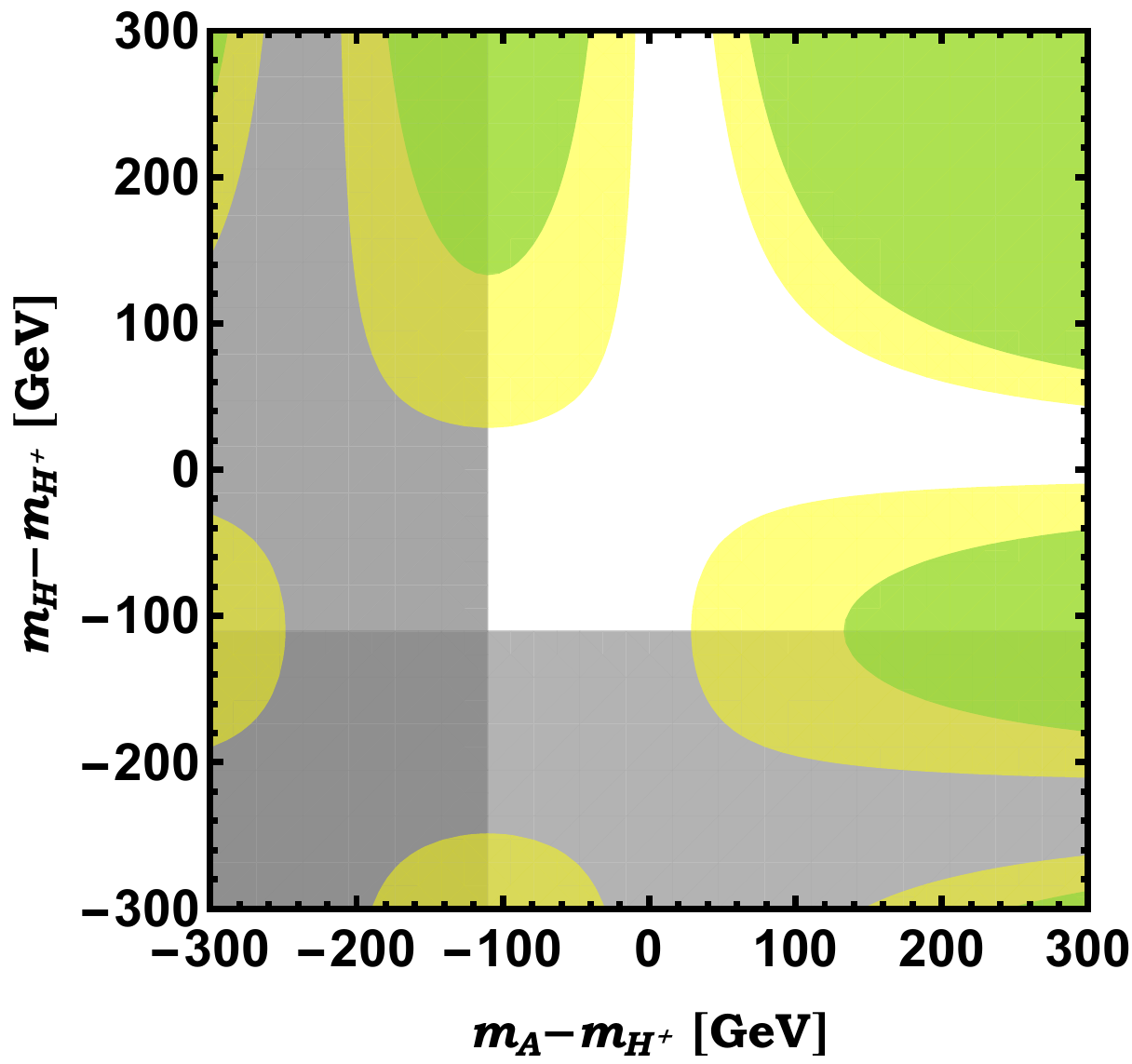}
\caption{
Scalar mass splittings allowed by the $T$ parameter constraint in the 2HDM.   The yellow and green shaded regions represent the 1$\sigma$ and 2$\sigma$ exclusion regions from the $T$ parameter constraint~\cite{Tanabashi:2018oca}. The horizontal and vertical grey shaded regions indicate the positivity criteria for $m_H >0$ and $m_A >0$, respectively. Here, we set  $m_{H^{\pm}}=110$ GeV.} \label{tpara}
\end{figure}

\begin{figure}[b!]
\includegraphics[width=0.28\textwidth]{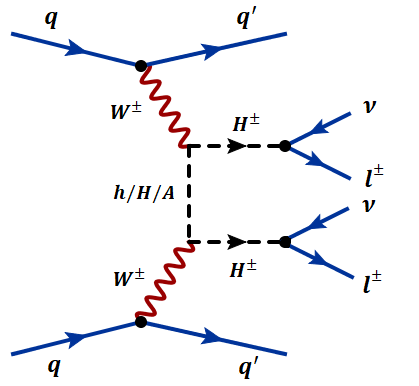}
\caption{Representative Feynman diagram for the signal $pp \rightarrow \tau^+\tau^+ jj +$ ${E\!\!\!\!/}_{T}$  at the LHC.} \label{lnv}
\end{figure}

\textbf{Electroweak precision constraints:}--
The effects of NP on the self-energies of the gauge bosons are parametrized in terms of oblique parameters $S, T,$ and $U$. From the EW precision data, these parameters impose strong constraints on any NP beyond the SM and  have been calculated at the one-loop level for general multi-Higgs-doublet models in Refs.  \cite{Funk:2011ad,Grimus:2007if,Grimus:2008nb,Babu:2018uik}.  In the alignment limit, the $T$ parameter in the 2HDM can be expressed as:
 \begin{align} 
    & T  =  \scriptstyle \dfrac{1}{16\pi s_W^2 M_W^2} \left\lbrace  { \mathcal{F}(m_{H^+}^2,m_{H}^2) + \mathcal{F}(m_{H^+}^2,m_{A}^2) 
    -\mathcal{F}(m_{H}^2,m_{A}^2) } \right\rbrace \,, \label{eq:T}
\end{align}
where the symmetric function $\mathcal{F}$ is given by
\begin{equation} \label{Fdef}
    \mathcal{F}(m_1^2,m_2^2) \  \equiv \  \frac{1}{2}(m_1^2+m_2^2) -\frac{m_1^2m_2^2}{m_1^2-m_2^2}\ln\left(\frac{m_1^2}{m_2^2}\right)\,.
\end{equation}
By analyzing these additional contributions, we find that the bound on the T parameter imposes strong restrictions on the mass splittings among the scalars in our scenario. As discussed above, in this work we set $m_{H^{\pm}}=110$ GeV, which is  consistent with the aforementioned LEP precision data.  We then turn on the mass splitting between charged scalar and the CP-even neutral scalar $H$ as well as between the charged scalar and the CP-odd neutral scalar $A$. Now, we investigate  the maximum possible mass splittings allowed by the $T$ parameter constraints. The corresponding region plot is shown in  Fig.~\ref{tpara}. The yellow and green shaded regions indicate the $1\sigma$ and 2$\sigma$ exclusion regions from the $T$ parameter constraint, respectively. The horizontal and vertical gray shaded regions corresponds to the positivity criteria for $m_H > 0$ and $m_A >0$, respectively. From this figure it is apparent that our scenario: $m^2_{H} \ll m^2_{H^{+}}=m^2_{A} \sim \mathcal{O}(110)$ GeV is well consistent with the EW precision constraints.

\textbf{Future implications at collider:}--
Here we discuss the testability of the proposed  scenario in the upcoming experiments. 
As we discussed earlier, explanations of the experimental data of $\Delta a_{e,\mu}$ solely depend on the existence of a light CP-even scalar.  This scenario can be tested at the LHC by looking at the novel process $pp \to H^\pm H^\pm jj \to \tau^\pm \tau^\pm j j + {E\!\!\!\!/}_{T}$, and the corresponding representative Feynman diagram is presented in Fig.~\ref{lnv}. It is interesting to note that if 
the mass splitting between the CP-even and CP-odd neutral scalars is turned off, then the amplitude for this process will be exactly zero. Correspondingly, our scenario will fail to explain the lepton AMMs, since a large mass splitting is essential to properly incorporate $\Delta a_{e,\mu}$ data as discussed above. Hence, observed deviations in the lepton AMMs are directly correlated with the  signal $pp\to \tau^\pm \tau^\pm j j +$ ${E\!\!\!\!/}_{T}$  in our set-up.  Due to this complementarity, this particular explanation of the electron and the muon $g-2$ within the 2HDM can be tested by this novel same sign charge lepton  process.   This same-sign charged lepton signature via vector-boson fusion process at the LHC has been studied extensively in Ref.  \cite{Aiko:2019mww}, although in a different context. We recast this analysis for our case and obtain the projected sensitivity for the signal   $pp \to H^\pm H^\pm jj \to \tau^\pm \tau^\pm j j +$ ${E\!\!\!\!/}_{T}$  at the LHC for centre of mass energy 14 TeV with integrated luminosity $\mathcal{L}=3$ ab$^{-1}$ and also for the centre of mass energy 27 TeV with integrated luminosity $\mathcal{L}=15$ ab$^{-1}$. We find that a charged scalar of mass 282 GeV (630 GeV) can be probed at the 14 TeV (27 TeV) LHC with integrated luminosity $\mathcal{L}=3$ ab$^{-1}$ ($\mathcal{L}=15$ ab$^{-1}$) while  there is 110 GeV mass splitting between $CP$-even ($H$) and  $CP-$ odd ($A$) scalar. These projected sensitivities are shown in Fig.~\ref{moneyplot} by black dashed lines.
Another interesting collider prospect that we briefly mention here is the decay of the SM Higgs into a pair of light scalars. As can be seen from Fig. \ref{moneyplot} that the light scalar of mass  $\sim \mathcal{O}$ (200) MeV, which is experimentally allowed can incorporate the deviations in the lepton AMMs. A light scalar around this mass region is particularly interesting since, the pair produced light scalars from the decay of the SM Higgs,  will further decay into two electrons or two muons. The process $pp \to h \to HH \to \mu^+ \mu^+ \mu^- \mu^-$, which is consistent with current experimental observations, however, this can be tested in future experiments, such as, HL-LHC and/or FCC-hh.  The associated detailed collider studies are beyond the scope of this paper and are left for  future work.  

\begin{widetext}
\begin{figure*}[th!]
\includegraphics[width=0.48\textwidth]{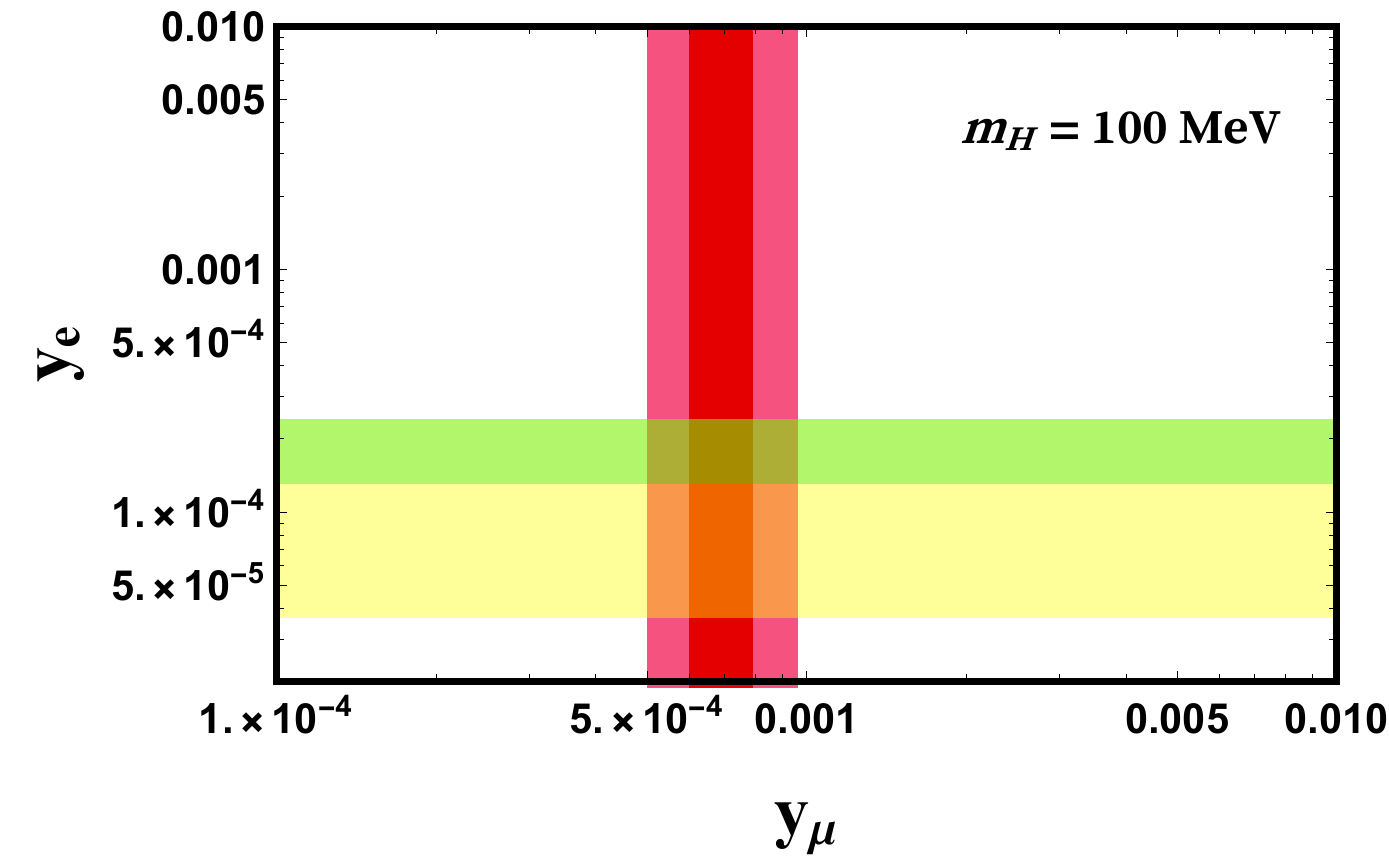}
\hspace{0.5cm}
\includegraphics[width=0.48\textwidth]{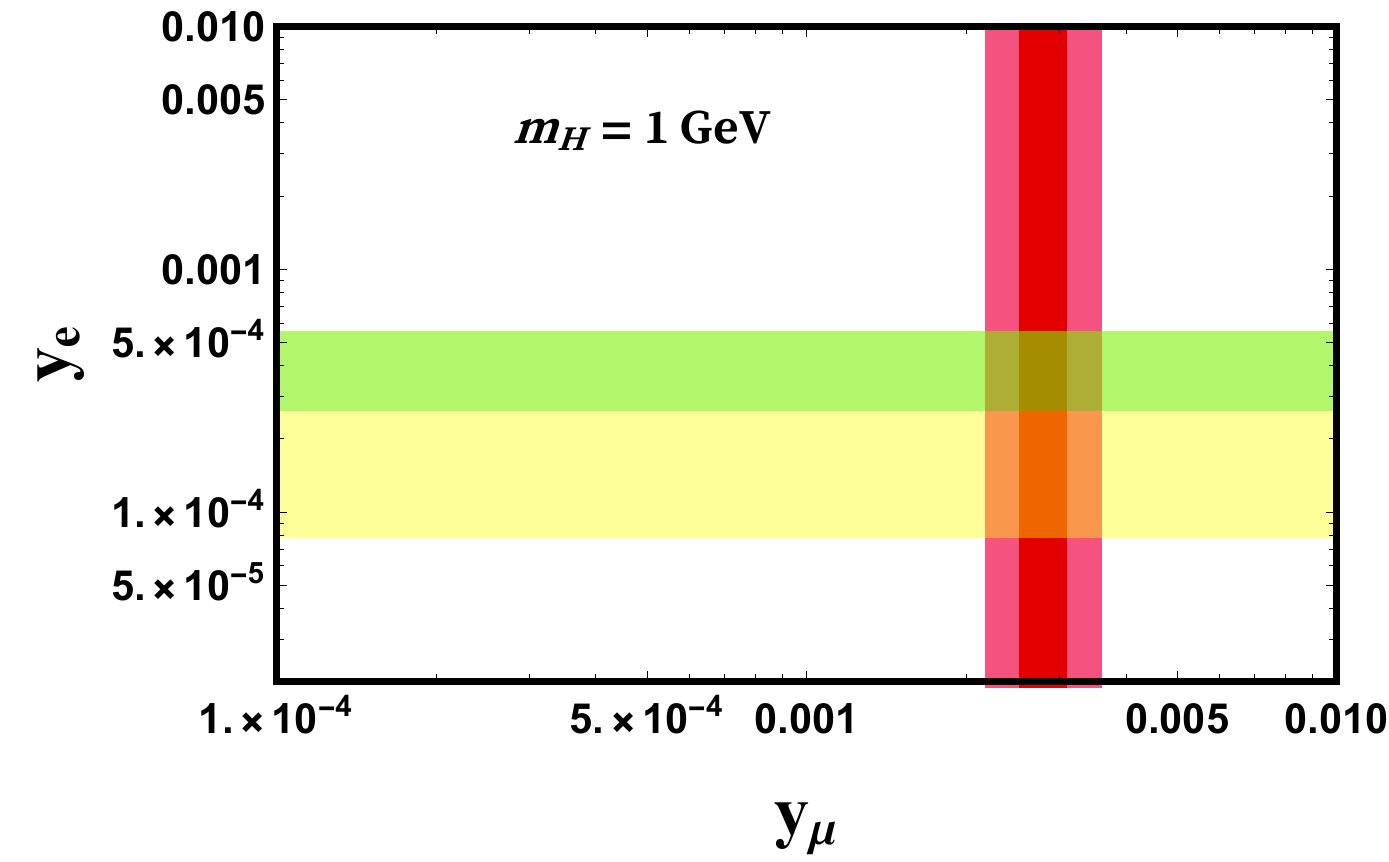}
\caption{The green (red) and yellow (pink) regions represent the experimental 1$\sigma$ and 2$\sigma$ bands for the electron (muon) AMM $\Delta a_e$ ($\Delta a_{\mu}$) in $y_e - y_\mu$ plane consistent with all the experimental constraints. We set $m_H=100$ MeV (1 GeV) for the left (right) panel. These plots demonstrate the required values of the electron and muon  Yukawa couplings for simultaneously explanations of $\Delta a_e$ and $\Delta a_\mu$ for a fixed value of $m_H$. It should be understood that as long as the Yukawa couplings are within the red (pink) and green (yellow) bands for the muon and the electron, respectively, a simultaneously solution to both $\Delta a_{e}$ and $\Delta a_{\mu}$ are achieved  within their $1\sigma$ ($2\sigma$) experimental measured values.
} \label{yukawaplane}
\end{figure*}
\end{widetext}

\textbf{Synopsis:}-- All the aforementioned current experimental  constraints applicable to our model, along with the future sensitivities are summarized in  Fig. \ref{moneyplot}. On top of that, we have also  plotted the experimentally measured values of AMMs of the electron and muon within 2$\sigma$ allowed range  that arise  from all the BSM degrees of freedom within this scenario.   It is evident  from this summary plot that despite numerous tight constraints, the CP-even scalar $H$ can remain light and live in the $\mathcal{O}(10)$-MeV to $\mathcal{O}(1)$-GeV mass range, and contribute simultaneously  to  both $\Delta a_{e,\mu}$ with correct magnitudes and signs. In the lower mass regime   $m_H<\mathcal{O}(10)$ MeV, it is incapable of explaining observed values of $\Delta a_{e}$ and $\Delta a_{\mu}$ simultaneously   regardless of other experimental constraints. On the contrary, in the opposite side of the parameter space, i.e., for  $m_H>\mathcal{O}(1)$ GeV, even though a concurrent explanation of both $\Delta a_{e}$ and $\Delta a_{\mu}$ is possible, however, various experimental constraints kill this portion of the parameter space.   Concerning these bounds depicted in  Fig. \ref{moneyplot}, a few comments are in order.   The $y_e$ coupling is independently constrained from electron beam-dump experiments \cite{Bjorken:1988as, Riordan:1987aw, Davier:1989wz}, the dark-photon searches through $e^+e^- \rightarrow \gamma H$ process at KLOE \cite{Anastasi:2015qla}, BaBar \cite{Lees:2014xha} and LEP~\cite{LEP:2003aa} experiment;  whereas the $y_\mu$ coupling is constrained  from the  $e^+e^- \rightarrow \mu^+\mu^- H$ searches at BaBar \cite{TheBABAR:2016rlg}  and LHC~\cite{Sirunyan:2018nnz} experiments. The $e^+e^- \rightarrow \mu^+\mu^- H$ searches at BaBar \cite{TheBABAR:2016rlg} and $e^+e^- \rightarrow \mu^+\mu^- $ searches at LEP \cite{LEP:2003aa} depend on both the Yukawa couplings $y_e$ and $y_\mu$. However, the constraints from $e^+e^- \rightarrow \mu^+\mu^- H$ is uniquely imposed  on $y_\mu$ since this process is mostly dictated by the $s-$ channel $Z/\gamma$ exchange  for somewhat smaller values of the Yukawa couplings relevant to our study. It is needless to mention that the process $e^+ e^- \to \mu^+ \mu^- H$ can be possible with the $H$,  emitted from $e^-$ or $e^+$ in the initial states. This process is solely dependent on the Yukawa coupling $y_e$ and which is already taken into account under the dark-photon searches through $e^+ e^- \to \gamma H$. Note that the $e^+e^- \rightarrow \mu^+\mu^- $ searches at LEP \cite{LEP:2003aa} can not impose bounds on $y_\mu$ independently. On the contrary,  bounds from $e^+e^- \rightarrow e^+e^- $ searches at LEP \cite{LEP:2003aa} are more stringent and over-shade the bounded parameter space from $e^+e^- \rightarrow \mu^+\mu^- $ searches for the region of our interest.

To make it more vivid, in Fig. \ref{yukawaplane}, we show the parameter space in the Yukawa coupling plane ($y_e - y_\mu$) which can explain the electron  (muon) AMM $\Delta a_e$ ($\Delta a_{\mu}$) consistent with all the experimental constraints considered in Fig. \ref{moneyplot} for two benchmark values corresponding to  $m_H=100$ MeV (left panel) and 1 GeV (right panel). As one can see from Fig. \ref{yukawaplane}, any values of $y_e$ within the green  and yellow bands can explain the the electron  AMM at 1$\sigma$ and 2$\sigma$ level, which is independent of the other Yukawa coupling  $y_\mu$. Similarly,  $y_\mu$ values within the red  and pink bands can satisfy the the muon  AMM at 1$\sigma$ and 2$\sigma$ level,  regardless of $y_e$. In these plots, we allow the values of $y_e$ and $ y_\mu$ such that the upper limits on the $y_e$ and $ y_\mu$ obey the stringent limits, mainly coming from the BABAR experiments.
It should be understood that as long as the Yukawa couplings are within the red (pink) and green (yellow) bands for the muon and the electron, respectively, a simultaneously solution to both $\Delta a_{e}$ and $\Delta a_{\mu}$ are achieved  within their $1\sigma$ ($2\sigma$) experimental measured values. Moreover, the  overlapping region does not carry any special significance.      

\section{Conclusion} \label{SECb}
Motivated by the recent precise measurement of the electron AMM $a_e$, which shows a significant deviation from the SM prediction, together with the intriguing deviation observed in the muon AMM $a_{\mu}$, we have proposed a novel scenario consisting of a light neutral scalar $H$ that is behind the origin of both these anomalies.  By properly taking into account theoretical and all existing experimental constraints, we have shown that a wide range of parameter space $\mathcal{O}(10)$ MeV $\leq m_H \leq$  $\mathcal{O}(1)$ GeV is still  allowed, which provides correct sizes and signs for both the $a_e$ and $a_{\mu}$. This is a highly non-trivial task since the light scalar is required  to have sizable couplings to all the SM charged leptons, and consequently is under severe experimental constraints. We have demonstrated how such a light CP-even scalar naturally arises from general 2HDM and serves the required purpose.  Our model predicts that the light scalar $H$ must be accompanied by nearly degenerate charged scalar $H^+$ and a pseudoscalar $A$ that have masses of  the order of the EW symmetry breaking scale. As we have shown by detailed analysis, the currently  allowed parameter space can be probed entirely in the upcoming experiments, which will either discover NP or completely rule out our scenario. A complementarity test of this scenario at the LHC  by seeking the novel process $pp \to H^\pm H^\pm jj \to l^\pm l^\pm j j + {E\!\!\!\!/}_{T}$ via same-sign pair production of charged Higgs bosons is also discussed.  

\begin{acknowledgments}
{\textbf {\textit {Acknowledgments.--}}} We thank  K. S. Babu and Carlos Wagner for useful discussions. The work of VPK was in part supported by US Department of Energy Grant Number DE-SC 0016013.
 \end{acknowledgments}

\bibliographystyle{utphys}
\bibliography{reference}

\end{document}